\documentclass[prb,showpacs,twocolumn,superscriptaddress]{revtex4}
\usepackage{amsmath}
\usepackage{graphicx}

\begin{document}
\title{The pattern of charge ordering in quasi-one-dimensional
organic charge transfer solids}
\author{R.T. Clay}\altaffiliation[Current address: ]
{Department of Physics and Astronomy,
Mississippi State University, Box 5167, Mississippi State, MS 39762}
\affiliation{ Department of Physics, University of Arizona
Tucson, AZ 85721}
\affiliation{Cooperative Excitation Project ERATO, Japan Science and
Technology Corporation (JST), Tucson, AZ 85721}
\author{S. Mazumdar}
\affiliation{ Department of Physics, University of Arizona
Tucson, AZ 85721}
\author{D.K. Campbell}
\affiliation{Departments of Electrical and Computer Engineering 
and Physics, Boston University, Boston, MA 02215}
\date{\today}
\begin{abstract}
We examine two recently proposed models of charge ordering (CO) in the
nominally $\frac{1}{4}$-filled, quasi-one-dimensional (1D) organic
charge transfer solids (CTS). The two models are characterized by site
charge density ``cartoons'' ...1010... and ...1100...,
respectively. We use the Peierls-extended Hubbard model to incorporate
both electron-electron (e-e) and electron-phonon (e-ph) interactions.
We first compare the results, for the purely electronic Hamiltonian,
of exact many-body calculations with those of Hartree-Fock (HF) mean
field theory. We find that HF gives qualitatively and quantitatively
incorrect values for the critical nearest-neighbor Coulomb repulsion
($V_c$) necessary for ...1010... order to become the ground state.
Second, we establish that spin-Peierls (SP) order can occur in either
the ...1100... and ...1010... states and calculate the phase diagram
including both on-site and intra-site e-ph interactions.  Third, we
discuss the expected temperature dependence of the CO and
metal-insulator (MI) transitions for both ...1010... and ...1100...
CO states. Finally, we show that experimental observations clearly
indicate the ...1100... CO in the 1:2 anionic CTS and the (TMTSF)$_2$X
materials, while the results for (TMTTF)$_2$X with narrower
one-electron bandwidths are more ambiguous, likely because the nearest
neighbor Coulomb interaction in these materials is near $V_c$.
\end{abstract}
\maketitle
\section{Introduction}
\label{intro_section}

Recent experiments
\cite{Sasaki95a,Pouget96a,Pouget97a,Kagoshima99a,Hiraki98a,Mazumdar99b,Hiraki99a,Meneghetti01a,Miyagawa97a,Biskup98a,Chow00a,Nad99a,Nad00a,Nad00b,Monceau01a,Miyagawa00a}
showing clear evidence for the existence of charge-order (CO) in
several 2:1 cationic organic charge transfer solids (CTS) have
stimulated considerable theoretical interest
\cite{Ung94a,Mazumdar99a,Mazumdar00a,Seo97a,Seo00a,Kobayashi97a,Kobayashi98a,Riera99a,Riera00a,McKenzie01a}.
As these materials are nominally $\frac{1}{4}$-filled (one electron or
hole per two sites) and involve electron-electron (e-e) interactions
(including nearest-neighbor Coulomb repulsion $V$), the apparently
obvious charge ordering in the quasi-1D CTS systems is the Wigner
crystal-like ``...1010...'' state, and numerous theoretical
\cite{Seo97a,McKenzie01a} and experimental
\cite{Hiraki98a,Nad99a,Nad00a,Nad00b,Monceau01a,Miyagawa00a} studies
have argued for this possibility.  The site occupancies `1' and `0'
actually correspond to 0.5 + $\epsilon$ and 0.5 - $\epsilon$,
respectively, with $\epsilon$ an important measurable quantity.  The
...1010... CO corresponds to a 4k$_F$ CDW (spatial period 2), where
k$_F = \pi/2a$ is the Fermi wavevector in the absence of Coulomb
interactions, with $a$ the lattice constant.  On the other hand, in
previous work we \cite{Ung94a,Mazumdar99a,Mazumdar00a} and others
\cite{Kobayashi97a,Kobayashi98a,Riera99a,Riera00a} have shown that for
realistic values of $V$ and when electron-phonon (e-ph) interactions
are included, a ``...1100...'' charge ordering can become the ground
state, and, importantly, that this CO state can explain the
``mysterious" states observed in several CTS
\cite{Sasaki95a,Pouget96a,Pouget97a}, in which coexisting charge and
spin density waves {\it of the same periodicity} occur.  The
...1100... CO corresponds to a 2k$_F$ CDW (spatial period 4) and has a
co-operative coexistence with the 2k$_F$ BOW or a mixed 2k$_F$ +
4k$_F$ BOW, which is why we termed it a ``Bond-Charge Density Wave''
(BCDW) \cite{Ung94a,Mazumdar99a,Mazumdar00a}.  In the present article,
we examine critically the theoretical and experimental evidence for
each of these two possible charge orderings in the quasi-1D organic
CTS.  Before proceeding to our analysis, it is useful to formulate in
a precise manner several key questions related to the charge ordering
in the organic CTS.  We focus here on four such questions.

First, the ...1010... CO ground state requires both strong on-site
Coulomb repulsion ($U >> t_0$ within the extended Hubbard model) {\it
and} nearest-neighbor Coulomb repulsion ($V$ within the extended
Hubbard model) to be greater than a critical value, $V_c(U)$.  For any
comparison to real materials determining the correct value $V_c(U)$ is
essential. Both Hartree-Fock (HF) and exact many-body methods have
been used to estimate $V_c(U)$. How reliable are the HF estimates
quantitatively, and do they exhibit the correct qualitative trends as
$U$ varies?

Second, charge ordering is only one of the phenomena observed in
organic CTS. At lower temperatures, broken symmetry states involving
spin ordering---spin-Peierls (SP) and spin density wave (SDW)
states---are typically observed, to say nothing of superconductivity.
Many of the quasi-1D CTS materials that show CO also feature SP ground
states at low temperature. Our previous work
\cite{Ung94a,Mazumdar99a,Mazumdar00a} establishing the BCDW state
proves that the 1100 CO {\it is} consistent with SP. To our knowledge,
none of the theoretical
\begin{figure}[tb]
\centerline{\resizebox{3.2in}{!}{\includegraphics{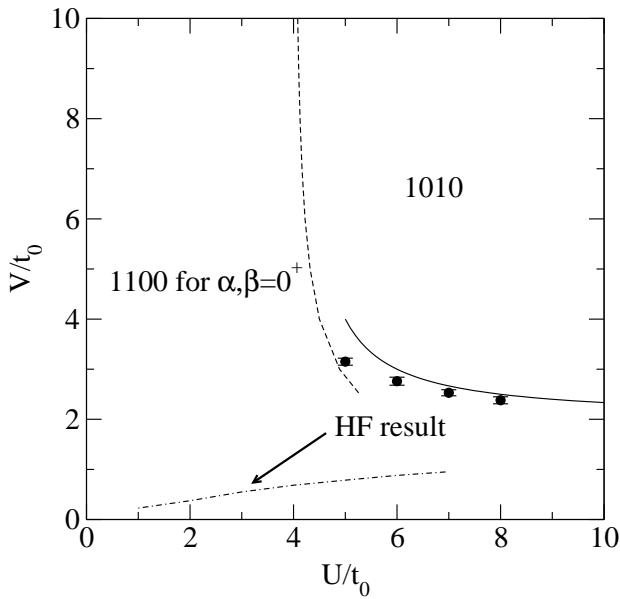}}}
\caption{Phase diagram of the 1D extended Hubbard model
at quarter-filling for positive values of $U$ and $V$. The 4k$_F$ CDW
exists only in the upper right hand corner, above the value of
$V_c(U)$ determined by the two strong coupling curves
(solid line: $U \to \infty$; dashed line: $V \to \infty$) and the quantum
Monte Carlo data points. Dashed-dotted line: Hartree-Fock data.}
\label{vcrit-fig}
\end{figure}
studies proposing the
existence of ...1010... CO \cite{Seo97a,McKenzie01a} has
investigated whether a SP phase can occur within the ...1010.... CDW
at lower temperatures. Indeed, an early quantum Monte Carlo study
\cite{Hirsch84a} can be taken to suggest that a SP phase could {\it
not} coexist with the ...1010... CDW. Is 1010 CO consistent with a SP phase 
at low temperatures, and if so, under what conditions?

Third, any successful theory of charge ordering within the organic CTS
must explain not only the mechanism and pattern for CO, but also the
temperature scale at which it occurs. In particular, any complete
theory of CO in these materials must be able to explain the existence
of different temperatures for the metal-insulator (MI), CO, and SP
transitions. Does either of the theoretical models for CO explain the
observed sequence of transitions?

Fourth, the ultimate arbiter in the ...1010... vs. ...1100... debate
is experiment, and in this regard it is clearly desirable to have as
broad a range of probes and materials as possible.  Both the 2:1
cationic CTS (with holes as carriers) and the 1:2 anionic CTS (with
electrons as carriers) have been studied extensively using similar
theoretical and experimental approaches. What is the experimental
evidence for ...1010... and for ...1100... in both these classes of
materials?

We address these questions in the remainder of this paper. In Section
\ref{vcrit} we introduce the the electronic extended Hubbard model and
compare exact to mean field calculations.  Here we establish that HF
theory exaggerates the role of the ...1010... CO and that the {\it
realistic} range of parameter values for which this CO is obtained is
rather narrow.  In Section \ref{4kfcdwsp} we introduce e-ph couplings
and show that it {\it is} possible for a spin Peierls transition to
occur with the ...1010... CO phase In Section \ref{temperature}, we
discuss the theoretical expectations for the sequence of temperatures
at which the various transitions--metal-insulator (MI), charge order
(CO) and spin-Peierls (SP)-- occur, establishing that within the
strictly 1D ...1100... CO model, $T_{MI} > T_{CO}=T_{SP}$, whereas
within the strictly 1D ...1010... CO model, $T_{MI} = T_{CO}>T_{SP}$.
In Section \ref{expt} we compare our theoretical results with
experimental data on a wide variety of CTS, both cationic and
anionic. These comparisons establish that in both the anionic 1:2 CTS
and the (TMTSF)$_2$X the dominant CO pattern appears to be ...1100...,
which in the latter case coexists with a SDW (because of
two-dimensional effects \cite{Mazumdar00a}) rather than a SP phase.
For the (TMTTF)$_2$X CTS, the situation appears less certain, but
...1010... order appears to be favored. We propose specific
experiments to distinguish definitively between the two different CO
patterns.  We also show in Section \ref{temperature} that the
experimentally observed temperatures of the MI, CO, and SP transitions
are not fully consistent with either ...1010...  or ...1100... models
in their strictly 1D forms.  Finally, in Section
\ref{conclusion_section}, we summarize our results and list some open
problems for future research.

\section{The 1D Extended Hubbard Hamiltonian}
\label{vcrit}

We first consider the 1D extended Hubbard Hamiltonian 
\begin{subequations}
\label{eqn-extHub}
\begin{eqnarray}
H &=& H_0 +H_{ee} \\
H_0 &=& -t_0 \sum_{j,\sigma}[c_{j,\sigma}^\dagger c_{j+1,\sigma}+
c_{j+1,\sigma}^\dagger c_{j,\sigma}] \\
H_{ee} &=& U\sum_{j}n_{j,\uparrow}n_{j,\downarrow} + 
V\sum_{j}n_{j}n_{j+1} 
\end{eqnarray}
\end{subequations}
In the above, $j$ is a site index,
$n_j=n_{j,\uparrow}+n_{j,\downarrow}$, and $\sigma$ is spin.  We focus
on the $\frac{1}{4}$-filled case, with the average number of electrons
(or holes) per site, $\rho=\frac{1}{2}$.  In addition to the purely
electronic terms in Eq.~(\ref{eqn-extHub}), we will also consider e-ph
couplings explicitly in section \ref{4kfcdwsp}.

That for $U \rightarrow \infty$ there exists a critical value ($V_c$)
of $V$ for the appearance of the 4k$_F$ CDW (...1010...) within the
$\frac{1}{4}$-filled 1D ``pure'' extended Hubbard model in
Eq.~(\ref{eqn-extHub}) has been known for decades \cite{Johnson72a},
and several detailed studies of $V_c$ for finite $U$ have appeared in
recent years \cite{Penc94a,Lin95a,Clay99a}. Nonetheless, it appears
that the implications of these theoretical results have been
insufficiently appreciated in the recent literature on organic
CTS. Thus, we believe these results and the underlying arguments bear
revisiting briefly here. In the limit $U \rightarrow \infty$, the
$\frac{1}{4}$-filled band of spin-$\frac{1}{2}$ electrons becomes
equivalent to a half-filled band of spinless fermions
\cite{Klein74a,Bernasconi75a}, which in the presence of $V$ can be
mapped using a Jordan-Wigner transformation to the Heisenberg XXZ
chain. This rigorous mapping and the exact solution for the XXZ model
establish that for $V>2t_0$, the system is in the (gapped)
``Ising-Heisenberg'' phase, which in terms of the original electronic
model corresponds to the 4k$_F$ CDW order ...1010... For $V<2t_0$, the
spin system is in the ``Heisenberg-XY'' phase and there is no gap,
which in the {\it pure} extended Hubbard model corresponds to a
Luttinger liquid (LL) with no charge order. Hence for $U \rightarrow
\infty$, the critical value of $V$ is $V_c(\infty) = 2t_0$.

For finite $U$, the value of $V_c$ at finite $U$ has been calculated
i) within strong coupling perturbation theory around $U \rightarrow
\infty$, where a second-order calculation yields \cite{Lin95a} $V_c(U)
\simeq 2t + \frac{2t^2}{U-4t} + \cdots$; ii) within strong-coupling
theory in the limit $V \rightarrow \infty$, where a similar
second-order calculation shows \cite{Penc94a,Lin95a} that for finite
$V$ the 4k$_F$ CDW phase boundary occurs at $U_c(V) \simeq 4t +
\frac{8t^3}{V^2} + \cdots$; and iii) for intermediate values of $U$
and $V$ by both exact diagonalization \cite{Penc94a} and quantum Monte
Carlo methods \cite{Lin95a,Clay99a}.  In Fig.~\ref{vcrit-fig} we show
the resulting boundary of the 4k$_F$ CDW phase in the $U,V$ plane; in
this figure, the solid and dashed curves represent the two strong
coupling expansions and the points with error bars represent our new
determination of the boundary at intermediate coupling using a quantum
Monte Carlo approach described in Ref. \onlinecite{Clay99a}.

\begin{figure}[tb]
\centerline{\resizebox{3.2in}{!}{\includegraphics{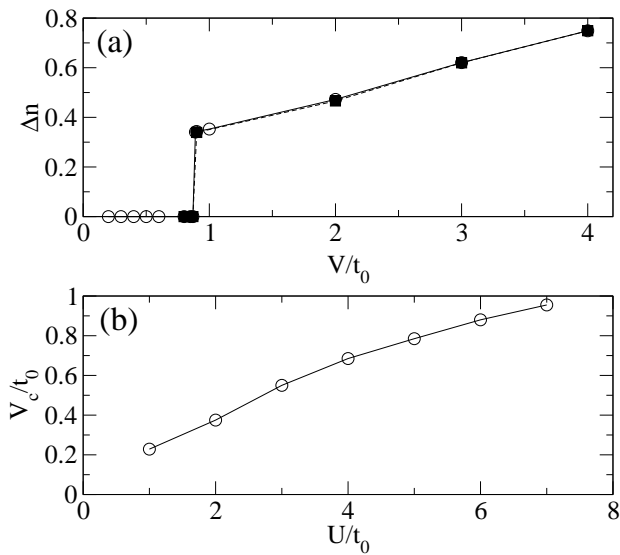}}}
\caption{(a) Charge disproportiation $\Delta n$ vs. $V$ within the HF
approximation for $U=6t_0$. Open (filled) symbols are for $N$=16 ($N$=32) 
rings. 
(b) $V_c$ within the HF approximation. Lines are guides to the eye. }
\label{vcrit-hf}
\end{figure}

The relevance of presenting the true phase boundary in the $U,V$ plane
(as determined by accurate many-body methods) becomes clear when one
considers the results obtained in a number of recent studies of the
$\frac{1}{4}$-filled 1D extended Hubbard model using mean-field
approximations \cite{Seo97a,Kobayashi97a,Kobayashi98a}. Using the HF
approximation for the $U$ interaction, and Hartree for the $V$
interaction, as is done in references \onlinecite{Seo97a}, leads to
drastically reduced values of $V_c$ compared to the true results: for
$U=5t_0$, the mean-field $V_c^{MF}=0.4t_0$ \cite{Seo97a}, as compared
to $V_c=(3.15 \pm 0.07) t_0$ expected from Fig.~\ref{vcrit-fig}.  In
Fig.~\ref{vcrit-hf}, we present results for our own unrestricted
HF calculations on finite size lattices.  In these calculations, both
$U$ and $V$ interactions were treated within the full HF
approximation, and no assumptions are made about the periodicity of
the solution.  The value of $V_c$ is readily determined by the point
at which $\Delta n$, the difference between charge densities on
adjacent sites, becomes nonzero (see Fig.~\ref{vcrit-hf}(a)).  As seen
in Fig.~\ref{vcrit-hf}(a), finite size effects are quite small, and
$V_c$ for the 64 site lattice did not differ significantly from 32 and
16 sites.  From Fig.~\ref{vcrit-hf}(b), we see that contrary to the
known exact results for the 1D extended Hubbard model, {\it within the HF
approximation} (i) the ...1010... CDW does occur for $U<4t_0$, and (ii)
the HF $V_c$ {\it increases} with $U$. For direct comparison of exact
and HF results, the data of Fig.~\ref{vcrit-hf}(b) are plotted as the
dashed-dotted line in Fig.~\ref{vcrit-fig} and clearly show that the
mean-field approximations incorrectly increase the parameter regime in
which ...1010... CO is expected in the extended Hubbard model.

\section{Charge and spin order including phonons in 1D}
\label{4kfcdwsp}

To answer the second question posed in the introduction, we explore
systematically the quantitative effects of adding e-ph interactions to
the extended Hubbard model in Eq.~(\ref{eqn-extHub}), resulting in a
``Peierls-extended Hubbard'' (PEH) model.  Our results establish
that a SP transition {\it can} occur within the ...1010... CO state
and clarify the regions of e-ph parameter space in which the two
competing forms of charge order exist.  Our PEH model follows from
replacing the one-electron part $H_0$ in Eq.~(\ref{eqn-extHub}) by
\begin{eqnarray}
H_0 &=& -\sum_{j,\sigma}[t_0-\alpha\Delta_j]B_{j,j+1,\sigma}
+ \beta\sum_{j}v_{j}n_{j} \nonumber \\
&+& \frac{K_1}{2}\sum_{j}\Delta_{j}^2 + \frac{K_2}{2} \sum_{j}v_{j}^2 
\label{eqn-phham}
\end{eqnarray}
In the above $\Delta_j=u_{j+1}-u_j$, where $u_j$ is the displacement
of the $jth$ atom from equilibrium. The amplitude of the internal
molecular vibration is given by $v_j$. $\alpha$ and $\beta$ are the
intersite and intrasite e-ph coupling constants, respectively, and
$K_1$ and $K_2$ are the corresponding spring constants. The kinetic
energy operator $B_{j,j+1,\sigma}=c^\dagger_{j+1,\sigma}c_{j,\sigma} +
c^\dagger_{j\sigma} c_{j+1,\sigma}$.  We determine (numerically)
self-consistent ground state solutions for the above PEH model as
functions of the parameters, measuring charge density $\langle n_j
\rangle$ and bond order $\langle \sum_\sigma B_{j,j+1,\sigma}
\rangle$.

We start with a summary of previous relevant results
\cite{Ung94a,Mazumdar99a,Mazumdar00a,Riera99a,Riera00a} concerning the
SP transition in the $\frac{1}{4}$-filled band.  These prior
investigations have established that for $V <V_c$ and for non-zero
$\alpha$ {\it or} $\beta$ (or both), the dominant broken symmetry
below the metal-insulator but above the insulator-insulator SP
transition is the dimerized 4k$_F$ BOW with uniform
\begin{figure}[tb]
\centerline{\resizebox{3.0in}{!}{\includegraphics{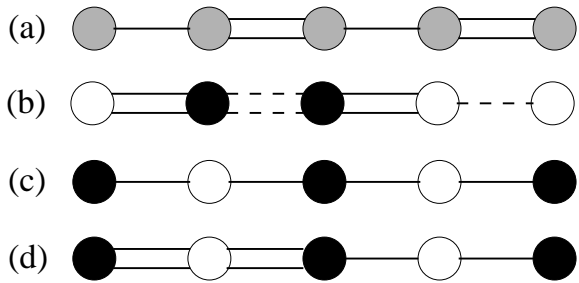}}}
\caption{The competing density wave ground states in the
Peierls-extended Hubbard Model at $\frac{1}{4}$-filling. Grey, black
and white circles correspond to site charges of 0.5, 0.5 + $\epsilon$
and 0.5 - $\epsilon$, respectively.  (a) The 4$k_F$ BOW state, with
dimerized bond orders $SWSW$ and uniform site charge densities. (b)
The BCDW state, with bond orders $SW'SW$ and the accompanying CO
...1100... (c) The 4$k_F$ CDW with ...1010...  CO and uniform bond
order. (d) The 4$k_F$ CDW-SP state, with ...1010... charge order and
bond orders $SSWW$. The charges on the ``unoccupied'' sites in (d) are
slightly different (see Fig.~\ref{slice}).}
\label{fig-dws}
\end{figure}
site charges (Fig.~\ref{fig-dws}(a)).  The SP transition here is a
{\it second} dimerization of the dimerized lattice, leading to the
BCDW state, which can be considered as a superposition of the 2k$_F$
and 4k$_F$ BOW's, and which is accompanied by the ...1100... CO
(Fig.~\ref{fig-dws}(b)).  The bond distortion pattern in the BCDW is
$SW'SW$, where $S$ is a strong bond, and $W$ and $W'$ are weaker bonds
with $W' > W$. In contrast, the bond distortions within the 4k$_F$
CDW-SP phase is of the form $SSWW$.  The competing CO is the
...1010... 4k$_F$ CDW, which has uniform bond orders at high
temperatures (Fig.~\ref{fig-dws}(c)).  The SP phase within the 4k$_F$
CDW phase, whose existence has not previously been studied or
established, would correspond to Fig.~\ref{fig-dws}(d). We demonstrate
below the existence and nature of this phase, which we hereafter refer
to as the 4k$_F$ CDW-SP phase. Within the 4k$_F$ CDW-SP phase there
occur alternate strong and weak bonds between the ``occupied'' sites
that are actually second neighbors (see Fig.~\ref{fig-dws}(d)). 
Previous work did not find this 4k$_F$ CDW-SP phase either because of
focusing primarily on $V<V_c$ \cite{Ung94a,Mazumdar99a,Mazumdar00a},
or because of including only one, but 
not {\it simultaneously both} of the e-ph
couplings $\alpha$ and $\beta$ \cite{Riera99a,Riera00a}. In this
section, we present new results for the $\frac{1}{4}$-filled band
for both $V<V_c$ and $V>V_c$, including both e-ph couplings $\alpha$
and $\beta$.

\begin{figure}[tb]
\centerline{\resizebox{3.4in}{!}{\includegraphics{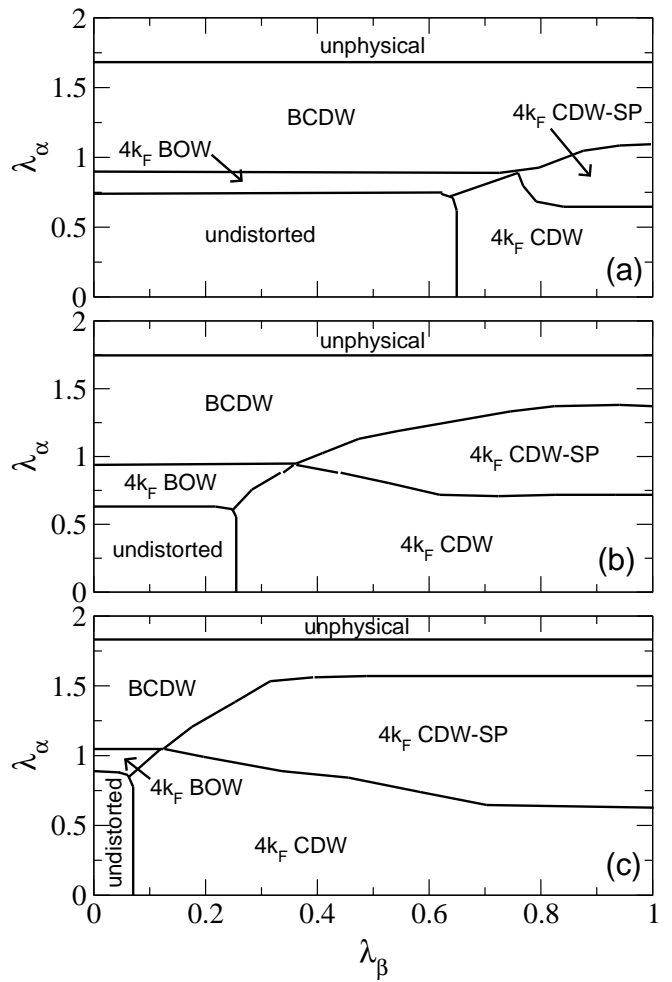}}}
\caption{The phase diagram in the $\lambda_{\alpha}-\lambda_{\beta}$
plane for $U=8t_0$ and (a) $V=2t_0$; (b) $V=3t_0$ ; and (c) $V=4t_0$
for a 16 site periodic ring. The boundaries between the the various
phases, drawn as continuous lines in the figures, are actually
determined numerically with an accuracy of $\sim 0.05$ in
$\lambda_{\alpha}$ and $\lambda_{\beta}$.  In the regions labeled
``unphysical'' the weakest bonds have negative bond orders due the
excessive lattice distortion that occurs for very large
$\lambda_{\alpha}$.}
\label{fig-phasediag}
\end{figure}

We have performed fully self-consistent exact diagonalizations of
finite-sized periodic rings.  The self-consistency equations derived
from $\frac{\partial \langle H\rangle}{\partial\Delta_j}=0$ and
$\frac{\partial \langle H\rangle}{\partial v_j}=0$ are
\begin{equation}
\Delta_j = -\frac{\alpha}{K_1} \langle B_{j,j+1,\sigma} \rangle,
\qquad v_j=-\frac{\beta}{K_2} \langle n_j \rangle 
\end{equation}
We present results in terms of dimensionless e-ph couplings
$\lambda_\alpha=\alpha^2/(K_1 t_0)$ and $\lambda_\beta=\beta^2/(K_2
t_0)$.  We have obtained results for systems of sizes $N$=8, 12, and
16.  In Fig.~\ref{fig-phasediag} we present a ground state phase
diagram ($N$=16) in the $\lambda_{\alpha} -\lambda_{\beta}$ plane, showing
regions in which the various broken symmetry states dominate.  In all
cases, $U=8t_0$; panels (a), (b) , and (c) correspond to $V=2t_0,
3t_0$, and $4t_0$, respectively.

Consider panel (a) with $V=2t_0$. Since $U =8t_0$, this value of $V$
lies below $V_c$ for the infinite system (see Fig.~\ref{vcrit-fig}).
For this 16-site system, as we increase $\lambda_{\alpha}$ from zero
(for fixed small $\lambda_{\beta}$), there is initially no
distortion. For $\lambda_{\alpha} \agt 0.7$, the ground state becomes
the 4$k_F$ BOW shown in Fig.~\ref{fig-dws}(a); while the site charge
densities remain uniform (n=0.5), the bond-order is now inhomogeneous
and has the form $SWSW$. For $\lambda_{\alpha} \agt 0.9$, the BCDW
discussed above and depicted in Fig.~\ref{fig-dws}(b) becomes the
ground state. Hence the bond-order pattern now is $SW'SW$ and the
charge order has the ...1100...  pattern. Next consider increasing
$\lambda_{\beta}$ from zero at fixed small $\lambda_{\alpha}$. Again
the state is initially undistorted, but when $\lambda_{\beta}$ reaches
a critical value, the 4$k_F$ CDW with ...1010... charge order and
uniform bond order becomes the ground state.  Importantly, we see that
a SP distortion {\it does} occur within the ...1010... charge-ordered
region, but the resulting 4$k_F$ CDW-SP state (shown in
Fig.~\ref{fig-dws}(d)) occurs only for fairly large values of both
$\lambda_\alpha$ and $\lambda_\beta$ when $U=8t_0$ and $V=2t_0$, at
least for this 16-site system. Finally, we note that in the region
labeled {\it unphysical} in Fig.~\ref{fig-phasediag} the value of
$\lambda_{\alpha}$ has become so large that self-consistency drives
the weakest hopping integral negative, rendering the results
unphysical in this context.

Panels (b) and (c) of Fig.~\ref{fig-phasediag} show several important
trends as $V$ is increased through and beyond the value of $V_c(8t_0)$
for an infinite system. First, the sizes of the undistorted, 4$k_F$
BOW and BCDW regions decrease considerably with increasing $V$, and
move to larger values of $\lambda_{\alpha}$, showing that it takes
stronger e-ph coupling to overcome the ``natural'' tendency toward
...1010... CO at large $V$. For the same reason, the sizes of the
4$k_F$ CDW and 4$k_F$ CDW-SP regions increase considerably with $V$
and occur for smaller values of $\lambda_{\beta}$. Second, the
existence of the 4$k_F$ CDW-SP ground state is very robust---$V$ does
not appear to suppress this SP order. 
Third, Fig.~\ref{fig-dws}
illustrates one clear qualitative difference between the two SP phases
{\it in this finite size system}. Provided that $\lambda_{\alpha}$ is
greater than a critical value depending on $V$
($\lambda_{\alpha_c}(V)$), the BCDW occurs for all $\lambda_{\beta}
\geq 0$. In contrast, occurrence of the 4$k_F$ CDW-SP state requires
both $\lambda_{\alpha}$ and $\lambda_{\beta}$ to be non-zero. Note
however that as $V$ is increased, the {\it minimum} value of
$\lambda_{\beta}$ (call it $\lambda_{\beta_c}(V)$), for which the
4$k_F$ CDW-SP state exists decreases rapidly towards zero, while the
value of $\lambda_{\alpha}$ necessary to produce this phase at
$\lambda_{\beta_c}(V)$ increases slightly.

These intriguing finite size results beg the question of what happens
to the system in the thermodynamic limit, $N \rightarrow\infty$. 
Unfortunately, our limited data do not allow us to perform a
reliable finite size scaling analysis, and we can thus provide only
tentative, partial answers at this stage. We believe that in the $N
\rightarrow \infty$ limit the undistorted and 4$k_F$ BOW phases
disappear; our evidence for this comes from our earlier studies of the
BCDW (hence for $V < V_c$) on long {\it open} chains \cite{Clay01a},
which showed that the center of these chains distorts naturally into
a BCDW but which showed no sign of a pure 4$k_F$ BOW. Similar studies for
$V>V_c$ have thus far proven inconclusive, and the $N \rightarrow
\infty$ behavior in this regime remains an important open question.

\begin{figure}[tb]
\centerline{\resizebox{3.2in}{!}{\includegraphics{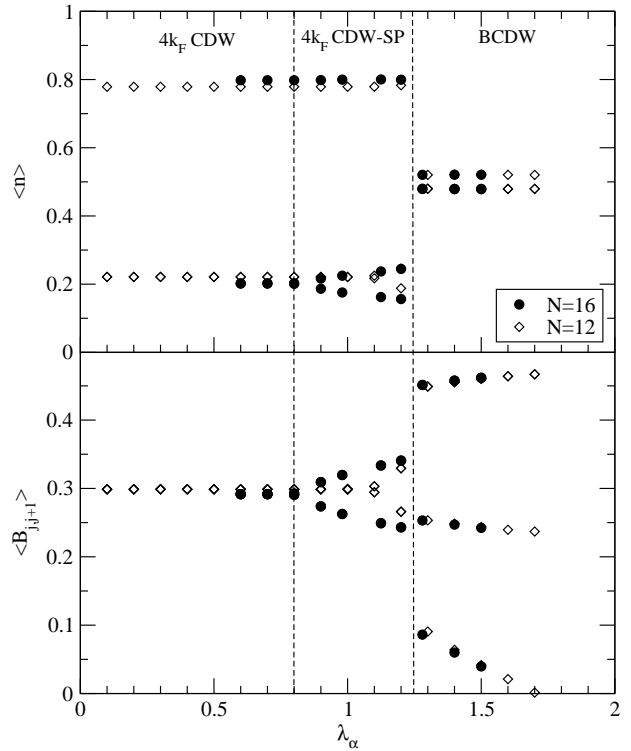}}}
\caption{Different possible values for charge densities and bond
orders for $U=8t_0$, $V=3t_0$, and $\lambda_\beta=0.6$ (see
Fig.~\ref{fig-phasediag}(b)).  Circles (diamonds) are data for $N$=16
($N$=12) site rings.}
\label{slice}
\end{figure}

Our self-consistent exact diagonalizations give results for site
charges, bond orders, and hopping integrals for all the phases,
homogeneous and inhomogeneous. 
In Fig.~\ref{slice} we present the actual site charge
densities and bond orders for $U=8t_0$, $V=3t_0$, and
$\lambda_\beta$=0.6.  This plot corresponds to a vertical ``slice''
through Fig.~\ref{fig-phasediag}(b), and shows the 4$k_F$-CDW,
4$k_F$-CDW-SP, and BCDW phases.  Several points are suggested by the
data in Fig.~\ref{slice}.
 
First, in the 4k$_F$ CDW-SP state, the sites with smaller charge
density are no longer equivalent---there is now (as expected from the
bond distortion) a small charge difference between these two
sites. Thus, the 4k$_F$ CDW-SP state is characterized by two different
bond orders and {\it three} different site charges, as opposed to the
two different charges and three different bond orders of the BCDW
state, which are also seen clearly in Fig.~\ref{slice}. We shall
return to this point momentarily. Second, the quantitative values of
the site charge differences, $\Delta n$, in the BCDW state are quite
small, whereas the $\Delta n$ in the 4$k_F$ CDW and 4$k_F$ CDW-SP
states are considerably larger. We have not been able to find, in our
finite system calculations, any BCDW state with large $\Delta n$, nor
any 4$k_F$ CDW or 4$k_F$ CDW-SP states will small $\Delta n$. Third,
and similarly, the bond order differences in the BCDW are always quite
large, whereas the bond order differences in the 4$k_F$ CDW-SP phase
are always quite small.  Again, these intriguing results for finite
size systems beg the question of the behavior for $N \rightarrow
\infty$, and again we cannot provide definitive statements because our
data are not sufficient to allow a reliable finite size scaling
analysis. One argument suggesting that the large $\Delta n$ may {\it
not} be an inevitable consequence of the ...1010... CO comes from the
$U \rightarrow \infty$ limit and the mapping to the Heisenberg-Ising
model; translating the exact results into fermion language shows that
charge difference in the ...1010... CO phase is given by $\Delta n
\sim \exp{-\frac{1}{(V-2t_0)}}$, so that it starts from zero and
remains small for a large region of $V > V_c=2t_0$
\cite{Johnson72a}. Clearly, further study of these potentially
significant differences in the $N \rightarrow \infty$ limit is
required.

Beyond the quantitative distinctions, however, there is the clear {\it
qualitative} distinction between the BCDW and the 4$k_F$ CDW-SP states: 
namely, the former always has two different site charges
and three different bond orders, whereas the latter has three
different site charges and two different bond orders. This qualitative
distinction will persist in the large system limit and is in principle
accessible to experimental techniques sensitive to the local charge or
bond order environment. For instance, if NMR measurements can be made
below the spin Peierls transition temperature ($T_{SP}$) in a given
CTS, they should be able to distinguish cleanly between the BCDW and
the 4$k_F$-CDW-SP states. We will return to this and related points in
our comparisons to experiment.

\section{Temperature dependence of CO transitions}
\label{temperature}

Our calculations in Sections \ref{4kfcdwsp} and \ref{vcrit}, as well
as virtually all other theoretical studies, have been limited to
ground state results only. However, the issue of the temperature
dependence of the CO and MI transitions is quite important in applying
the results of 1D model calculations to experimental systems. In this
Section, we discuss the temperature dependence expected for
...1010... and ...1100...  models of CO.

\begin{figure}[tb]
\centerline{\resizebox{3.2in}{!}{\includegraphics{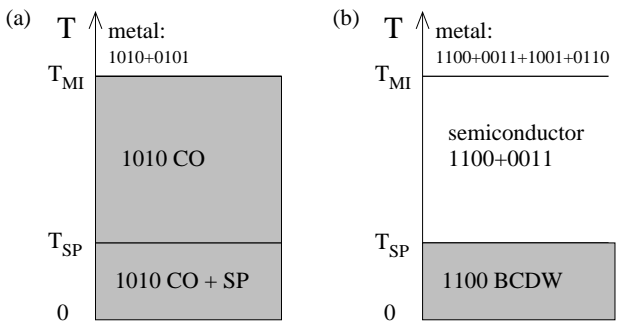}}}
\caption{Schematic showing expected temperature dependence of (a)
...1010... and (b) ...1100... CO models.  $T_{SP}$ and $T_{MI}$
indicate SP and MI transition temperatures respectively. The shaded
region indicates temperature range where CO is found.}
\label{fig-temp}
\end{figure}

As the ...1010... CO is driven by $V$, it is expected to give CO
already at quite high temperatures.  The series of transitions can be
understood by recognizing that the ...1010... CO has two equivalent
configurations, ...1010... and ...0101..., corresponding to a
double-well potential.  For temperatures above the CO transition
temperature, both possible configurations ...1010... and
...0101... have equal weight in the partition function, resulting in
no net CO and a metallic state.  The CO transition then corresponds to
a symmetry breaking between ...1010... and ...0101..., with the result
that CO and MI transitions (at $T_{CO}$ and $T_{MI}$ respectively)
happen at the {\it same} temperature.  As we have shown in Section
\ref{4kfcdwsp}, a SP distortion {\it can occur} within the
...1010... CO phase, but its scale is set by the e-ph couplings and is
expected to be well below the electronically driven
$T_{MI}=T_{CO}$. Thus within the ...1010... model (see
Fig.~\ref{fig-temp}(a)), one expects two transitions, a combined high
temperature CO/MI transition
followed by a low temperature SP transition.

In the ...1100... CO model, one must consider a quadruple-well
potential \cite{Mazumdar00a}. At the highest temperatures there are
four equivalent configurations, ...1100..., ...0011..., ...0110...,
and ...1001..., again giving a metallic state when all four have equal
weights. The high temperature transition corresponds to a symmetry
breaking that leaves ...1100... and ...0011... with equal weight,
resulting in no net CO, but an insulating state. Then at lower
temperature, the ...1100.../...0011... symmetry breaks, resulting in
CO.  As the ...1100... CO is a {\it cooperative} BCDW state, CO and
bond distortion occurs {\it simultaneously} \cite{Mazumdar00a}.  Thus
within the ...1100... model (see Fig.~\ref{fig-temp}(b)), one again
expects two transitions, a high temperature MI transition, followed by
a low temperature combined CO/SP transition.

\section{Interpretation of experiments}
\label{expt}

If detailed comparisons of models to real systems are to be attempted,
accurate estimates of the e-e and e-ph parameters are required.
Determination of $U$ and $V$ from quantum chemistry calculations is
difficult, and in general such calculations give values that are too
large \cite{Castet96a}. Fortunately, there is by now general agreement
on the magnitudes of $t_0$ in the different materials, and reasonable
values of $U$ and $V$ can be estimated from the experiments.  We
consider primarily materials based on the TCNQ, TMTSF, and TMTSF
molecules.  It is generally accepted that the values of $t_0$ in the
cationic TMTTF/TMTSF and anionic TCNQ CTS lie between 0.1 -- 0.25 eV.
We then agree with previous investigators that $U/t_0$ in (TMTTF)$_2$X
ranges from 7 - 12 \cite{Castet96a,Mila95a} and is less than 6 in
(TMTSF)$_2$X \cite{Mila95a,Jacobsen86a}.  Based on comparing
relative 4k$_F$ responses\cite{Mazumdar83a,Pouget88a,Kagoshima83a} (a
signature of large $U/t_0$), one may conclude that for TCNQ systems
$U/t_0$ is intermediate between the TMTTF and TMTSF limits.  There are
far fewer estimates of $V/t_0$: values of 2.8 for (TMTTF)$_2$X and 2.0
for (TMTSF)$_2$X have been proposed by Mila \cite{Mila95a}. We note
that whether or not $V > V_c(U)$ can be determined from the
experimental pattern of the CO, if it is known. On the other hand,
given the relatively small $U/t_0$ in (TMTSF)$_2$X, it is unlikely
that $V > V_c(U)$ in this system (recalling the additional condition
that $V < U/2$, based on general considerations of the nature of
realistic Coulomb interactions.).  To summarize, in the quasi-1D
materials, we believe that $(U/t_0)^{TMTTF} > (U/t_0)^{TCNQ} >
(U/t_0)^{TMTSF}$.  For the e-ph couplings, the dimensionless coupling
constants $\alpha^2/K_1t_0$ and $\beta/K_2t_0$ are even more difficult
to estimate \cite{Pedron94a,Pedron95a,Meneghetti96a}.  For our overall
purposes, it suffices to note that there is evidence for both nonzero
$\alpha$ and $\beta$.  Finally, although in this paper we do not
explicitly model the effects of transverse coupling of the 1D chains,
(see reference \onlinecite{Mazumdar00a} for a quantitative discussion
of these effects), we shall comment on these higher dimensional effects in
some of the experimental discussion below.  In this regard it is
important to recall that the electronic anisotropy and effective e-e
interaction strengths can vary {\it independently} among the CTS.  The
relative strength of the e-e interactions is measured by the ratios
$U/t_0$ and $V/t_0$, whereas the effective dimensionality is measured
by $t_{\perp}/t_0$). These three ratios vary considerably among the
materials. In particular, the TMTTF materials are both highly 1D and
very strongly correlated, whereas the TMTSF materials are less 1D and
also less strongly correlated.
 
Let us turn now to discuss the
experimental situations with individual classes of materials.

\subsection{1:2 anionic CTS}

As mentioned in the Introduction, we believe that a complete
understanding of the CO phenomenon in the $\frac{1}{4}$-filled band
CTS requires that both cationic and anionic CTS be examined on the
same footing. The availability of large single crystals in the case of
the anionic systems makes these particularly attractive.  On the other
hand, the anionic CTS are much older and many different contradictory
claims exist in the literature. We therefore focus only on the
materials for which clear evidence for CO has been found. The systems
we consider are MEM(TCNQ)$_2$, TEA(TCNQ)$_2$, (DMe-DCNQI)$_2$Ag, and
(DI-DCNQI)$_2$Ag.  Preliminary discussions of the CO in MEM(TCNQ)$_2$
and TEA(TCNQ)$_2$ were also given in reference \onlinecite{Ung94a}.

The most direct evidence for the ...1100... CO has been found in
MEM(TCNQ)$_2$ and TEA(TCNQ)$_2$.  The SP transition in MEM(TCNQ)$_2$
occurs at 17.4 K, and a very careful neutron diffraction study was
performed for deuterated samples at 6 K \cite{Visser83a}.  This work
clearly established the $SW'SW$ bond distortion pattern associated
with the ...1100... BCDW for T $<$ T$_{SP}$ in MEM(TCNQ)$_2$.
Similarly, the charge and bond modulation pattern for TEA(TCNQ)$_2$
have been investigated by X-ray
\cite{Kobayashi70a,Filhol84a,Farges85a,Farges85b} and neutron
\cite{Filhol80a} measurements.  Below the structural transition at
210K, the amplitude of the $SW'SW$ distortion increases rapidly, and
...1100...  CO was found \cite{Farges85a,Kobayashi70a}.

The pattern of the CO in (DI-DCNQI)$_2$Ag (but not in
(DMe-DCNQI)$_2$Ag), was until recently controversial.  The
(DI-DCNQI)$_2$Ag system was originally claimed to show the
...1010... CO from NMR experiments \cite{Hiraki98a}. It was pointed
out by the present authors that some of the experimental results,
especially the appearance of the CO at a temperature lower than the
metal-insulator transition (see Section \ref{temperature}), perhaps
indicated the ...1100... order \cite{Mazumdar99b}.  Initial optical
measurements had detected evidence for a 4k$_F$ BOW without CO already
at room temperature, which had suggested that a ...1100... CO
accompanies the second lattice distortion that occurs at lower
temperature \cite{Meneghetti01a}.  A first X-ray structural study
found moderately strong satellite intensities in the 4k$_F$ phase
(10$^{-3}$ of the Bragg reflection), accompanied by high $l$ indices
\cite{Nogami99a}.  This was interpreted to indicate a displacive
transition \cite{Nogami99a}, suggesting (though not necessarily
proving) that the 4k$_F$ phase is a BOW and not a CDW. This would
again suggest that the CO that occurs at low temperature is the
...1100... CO.  A later X-ray experiment included both
(DI-DCNQI)$_2$Ag and the structurally related material
(DMe-DCNQI)$_2$Ag \cite{Nogami99b}. The strong 4k$_F$ reflection in
the latter material (10$^{-3}$ -- 10$^{-2}$ of the Bragg reflection),
along with the occurrence of 2k$_F$ reflections, indicates that the
4k$_F$ phase is a BOW (as agreed upon also by the authors of
Ref.~\onlinecite{Nogami99b}).  
This would again indicate the ...1100... CO in the 2k$_F$ phase.  No
2k$_F$ reflections were found in (DI-DCNQI)$_2$Ag, and the index
dependence of the 4k$_F$ reflections were also different from that in
(DMe-DCNQI)$_2$Ag \cite{Nogami99b}.  Based on this, the authors of
Ref.~\onlinecite{Nogami99b}
suggested that unlike in (DMe-DCNQI)$_2$Ag, the 4k$_F$ order in
(DI-DCNQI)$_2$Ag is a CDW, although their calculated satellite
intensities, with the assumption of a ...1010... CO accompanied by
molecular distortions, were still lower than that of the observed
intensities. The occurrence of a 4k$_F$ BOW in (DMe-DCNQI)$_2$Ag and a
4k$_F$ CDW in (DI-DCNQI)$_2$Ag, though not impossible, would have been
rather mysterious.  This mystery has very recently been resolved by
Menghetti and collaborators \cite{Meneghetti02a}. The authors have
analyzed the temperature-dependent vibronic and vibrational infrared
absorptions in the (DI-DCNQI)$_2$Ag system in great detail and have
concluded that the low temperature phase is a 2k$_F$ bond tetramerized
phase with the ...1100... CO pattern. Indeed, the authors also find
that $\Delta$ n in this system is rather small, in agreement with the
predictions of our finite size calculations in Section~\ref{4kfcdwsp}.

To conclude, the CO pattern in all 1:2 anionic CTS for which detailed
experimental data are available is ...1100....

\subsection{(TMTTF)$_2$Br and (TMTSF)$_2$X}
(TMTTF)$_2$Br exhibits behavior similar to that of (TMTSF)$_2$X, in
that the low temperature phase here is the mixed CDW-SDW
\cite{Pouget96a,Pouget97a} and different from the other TMTTF-based
materials.  We therefore discuss this material together with the
(TMTSF)$_2$X.  There is strong evidence for the ...1100... CO in this
class of materials, as we now discuss.

First, we have pointed out in the above that $U/t_0$ and $V/t_0$ in
TMTSF are very likely smaller than that in the TCNQ solids.  Given the
clear evidence for the ...1100... CO in the 1:2 TCNQ solids, the
occurrence of the ...1010... CO in TMTSF therefore appears
questionable.  Specifically, previous parameterizations of
(TMTSF)$_2$X have all put $U/t_0 \leq 6$, which makes $V_c/t_0$ $\sim$
2.8.  Taken together with the condition $V < U/2$, this makes the
...1010.... order very unlikely.

Direct evidence for the ...1100... order in (TMTSF)$_2$X and 
(TMTTF)$_2$Br
comes from the determination that (a) there exists a coexisting
CDW-SDW in these systems below T$_{SDW}$, and (b) the periodicities of
these density waves are both 2k$_F$.
\cite{Pouget96a,Pouget97a}. Since the ...1010... CO would give a
4k$_F$ charge periodicity, this can then immediately be ruled out.
The coexisting CDW-SDW has been demonstrated in our previous quasi-2D
calculation within the Hamiltonian of Eq.~(\ref{eqn-phham})
\cite{Mazumdar99a}, where we demonstrated that for small but nonzero
interchain coupling the proper description of the ground state is a
bond-charge-spin density wave (BCSDW), with the CO being ...1100...

\subsection{(TMTTF)$_2$X}
In a series of recent papers, Nad et al. have found evidence for CO in
various (TMTTF)$_2$X materials, from dielectric permittivity studies
\cite{Nad99a,Nad00a,Nad00b,Monceau01a}. Importantly, the CO is found
both with centrosymmetric anions X = PF$_6$, AsF$_6$, SbF$_6$ etc. and
with the asymmetric anion ReO$_4$.  Observation of ferroelectric
behavior \cite{Monceau01a} suggests removal of all symmetry elements,
and based upon this it has been suggested that the CO pattern here is
of the ...1010... type \cite{Monceau01a,Riera01a} and is accompanied
by anion motion.  Independent and direct verifications of CO in X =
PF$_6$ and AsF$_6$ have come from NMR studies \cite{Chow00a}.
However, while the presence of some form of CO in (TMTTF)$_2$X is
undisputed, we believe there is not yet enough evidence to say
definitively whether the pattern of CO is ...1010... or
...1100... Below we review the evidence from several different perspectives 
and suggest a
possible resolution.

(i) The very high temperatures at which the CO is observed in some of
these CTS (T$_{CO}$ = 154 K and 227.5 K in X = SbF$_6$ and ReO$_4$,
respectively) indeed suggests the ...1010... order, since long range
...1100... order \cite{Ung94a,Mazumdar99a,Mazumdar00a} is associated
with the SP transition, which usually occurs at much lower
temperatures.
One possibility (not yet thoroughly explored, let alone established)
that would allow higher temperature CO within the ...1100... scenario
is that in at least some of the (TMTTF)$_2$X the CO is associated with
pre-transitional structural fluctuations from the SP
transition. Theoretically, it has been claimed that such fluctuations
of the SP order may become visible at temperatures as high as 3 -- 4
T$_{SP}$, where T$_{SP}$ is the SP transition temperature
\cite{Schulz87a}.  Experimentally, structural fluctuations associated
with the SP transitions have been observed in the related SP materials
(BCPTTF)$_2$PF$_6$ and (BCPTTF)$_2$AsF$_6$ at temperatures as high as
100 K and 120 K, respectively, even though their SP transition
temperatures are T$_{SP}$ = 37K and 34 K, respectively
\cite{Liu93a,Dumoulin96a}.  It is then intriguing that in
(TMTTF)$_2$PF$_6$ scattering at 2k$_F$ in X-ray experiments begins to
appear at nearly the same temperature \cite{Pouget97a} at which CO
appears in NMR \cite{Chow00a}.

(ii) NMR experiments lead to estimates of a lower limit of 0.5 for $\Delta n$
\cite{Zamborszky02a}, which also supports the ...1010... CO, since
within our finite size calculations we have never found such a large
$\Delta n$ within the 1D ...1100... phase. While the large $\Delta n$
would be in agreement with the ...1010... order, the difficult of
determining $\Delta n$ unambiguously from these experiments is shown
by the results similar NMR experiments on the inorganic material
NaV$_2$O$_5$, in which also CO involving the vanadium sites is
observed and in which the pattern of the CO remains controversial
after considerable study. These experiments have not led to $\Delta n$
values that are consistent: widely different $\Delta n$, ranging
anywhere from 0.1 to 0.8 have been reported in the literature in this
case \cite{Revurat00a,Ohama00a,Nakao00a}.

(iii) Recent NMR experiments on X=AsF$_6$ under pressure also find
that the CO is very sensitive to pressure and is suppressed for
pressures greater than $\sim$0.15 GPa \cite{Zamborszky02a}. The great
sensitivity to pressure raises further questions regarding the CO in
(TMTTF)$_2$X: if the CO in (TMTTF)$_2$X is of the ...1010...  pattern
and hence driven by large $V$, one would expect that increased
$t_\perp/t_0$ from the application of pressure would have little
effect on the CO. Importantly, studies of two-leg $\frac{1}{4}$-filled
ladders find $V_c(U)$ in the ladder to be quite similar to the value
in 1D for small $t_\perp/t_0$ \cite{Vojta01a}.

(iv) It has been suggested that the MI transition at higher
temperatures is not driven by intrastack 4k$_F$ bond dimerization
(which is unaccompanied by CO), but by the anion potential
\cite{Monceau01a}.  Ascribing the metal-insulator transition to the
anion potential, however, is also problematic.  For instance, if the
anion potential leads to a 4k$_F$ bond dimerization, there is further
increase in $V_c(U)$, and the ...1010... CO becomes even less likely
\cite{Shibata01a}.  On the other hand, if the anion potential led to
the ...1010... CO directly, evidence for the CO should have been found
already at the metal-insulator transition.  As regards the widely
differing T$_{CO}$, it is possible that different anion sizes lead to
different interchain couplings, which in turn affect the $V_c(U)$.

(v) Finally, why is T$_{CO}$ is so {\it low} in some of these systems
(T$_{CO} \sim$ 65 K and 100 K in X = PF$_6$ and AsF$_6$, respectively
\cite{Chow00a})?  The charge localization/metal-insulator transition
$T_{MI}$ in (TMTTF)$_2$PF$_6$ occurs at $>$ 200 K {\it above} the CO
temperature, while (as explained in Section \ref{temperature}) for the
...1010...  model one expects $T_{CO}=T_{MI}$.

A possible resolution to (i)---(v) is that the higher temperature CO
found in (TMTTF)$_2$X is indeed ...1010..., while the lower
temperature SP state has ...1100... CO.  Transitions at finite
temperature are determined not by the ground state energy but by the
free energy.  At temperatures T $\leq$ T$_{MI}$, the excitations of a
strongly correlated system are predominantly spin excitations. Because
of the greater multiplicities of high spin states, the free energy at
high T (but below T$_{MI}$) is then dominated by high spin states.
Consider now the fully ferromagnetic state, which is clearly described
within the {\it spinless} fermion Hamiltonian limit of
Eq.~(\ref{eqn-extHub}). Thus $V_c$ in this state remains 2$t_0$ even
for finite $U$ and is smaller than the $V_c$ for the ground state with
total spin S = 0. For similar reasons we expect for finite $U$, $V_c$
to increase progressively as the total spin S decreases.  Therefore,
for a system in which $V$ is slightly lower than $V_c$ (S = 0) but
higher than $V_c$ (S = S$_{max}$) we see that at high temperatures,
where high spin excitations are thermally accessible, the CO pattern
can be ...1010... At lower temperatures where the high spin states
become thermally inaccessible and the free energy is dominated by low
spin states there can be a switching to the ...1100... CO. We are
current exploring this scenario in detail.

Finally, we suggest two experiments that can settle the issue of CO in
(TMTTF)$_2$X. First, X-ray or neutron diffraction experiments can
probe the actual bond distortions below the SP transition: with
sufficient sensitivity, they should be able to distinguish the $SW'SW$
distortions associated with the ...1100... CO from the $SSWW$
distortions associated with the ...1010... CO.  Second, as shown in
Section \ref{4kfcdwsp}, we predict three different charges below
T$_{SP}$ for the case of the ...1010... CO (rather than the two
charges associated with the ...1100... state). If NMR experiments
can be extended to this temperature region, further splitting of the
NMR lines is predicted. Unfortunately, our results also suggest that the
charge difference between the two `0's is rather small, and an
extremely sensitive probe might be necessary to determine the new
splitting.

\section{Conclusion}
\label{conclusion_section}

Our conclusions can be summarized as follows. First, formation of the
Wigner crystal-like ...1010... CO requires V within the 1D extended
Hubbard model that is much larger than that predicted within mean
field theory. Further, mean field theory predicts the incorrect
behavior of $V_c(U)$ as a function of $U$. Accurate many-body
calculations show that for finite $U$, $V_c(U) > 2|t_0|$ and increases
with decreasing $U$.

Second, we have given a complete theory of the SP transition that can
occur in the correlated 1D $\frac{1}{4}$-filled band. 
The SP phase with the ...1100... CO is a BCDW, with two different
charges and three different bonds, the bond distortion pattern being
$SW'SW$. In contrast, the SP phase in the ...1010... CO has two kinds
of bonds and three distinct charges; the bond distortion pattern here
is $SSWW$, and this distortion makes the charges on the sites labeled
`0' unequal. 
Several distinct types of experiments have confirmed the
existence of two kinds of charges and three kinds of bonds in a number
of 1:2 anionic CTS.
In (TMTSF)$_2$X, experiments find coexisting CDW and SDW with the
{\it same} periodicity, which as discussed above precludes the 
...1010... CO.
The situation in the
(TMTTF)$_2$X is less clear, as experiments showing the existence of CO
do not directly measure the specific CO pattern. We know of no
experiments to date that have confirmed the existence of the three
types of charges and two types of bonds associated with the SP phase
of the ...1010...  We believe that the NMR and X-ray experiments
that could confirm this should have high priority.

Third, within the strictly 1D models, the two charge orders predict
different scenarios for the temperature dependence of the CO. For the
...1100... CO, $T_{MI} > T_{CO} = T_{SP}$, whereas for the
...1010... C0, $T_{MI} = T_{CO} > T_{SP}$. Neither of these results is
in agreement with all the experiments, which in general show $T_{MI} >
T_{CO} >T_{SP}$.  We believe that weakly 2D inter-chain couplings,
which explain the transition from CDW-SP to CDW-SDW ground states in
some of the materials \cite{Mazumdar00a}, may also be the key to
understanding the ordering of $T_{MI}$, $T_{CO}$, and $T_{SP}$ in the
different CTS. Support for this belief comes from recent results
on the strongly 2D $\theta$-(BEDT-TTF)$_2$X materials,
in which $T_{MI}$ occurs simultaneously with $T_{CO}$, followed by a
lower temperature spin-gap transition \cite{Mori98a,Mori98b}. This
sequence of transition temperatures, as well as the 2D CO and spin-gap
formation, can be understood within a 2D, six-fold coordinated lattice
model with dominant ...1100... CO order \cite{Clay02a}.

Fourth, our calculations suggest several problems for further
study. Apart from the experiments seeking to confirm a ``three-charge,
two bond'' 4$k_F$-CDW-SP state, these include theoretical studies providing a
full explanation of the sequence of MI, CO, and SP transitions as the
temperature is decreased, an accurate determination of the phase
diagram in the e-ph coupling constant plane in the thermodynamic
limit, and a quantitative understanding of the role of broken
symmetries associated with higher spin states at finite
temperature. We are currently investigating these theoretical issues.

Finally, during preparation of this paper we became aware of the
recent work of Shibata et al. \cite{Shibata01a}, who investigated the
purely electronic extended Hubbard model in 1D. Although they did not
discuss the ...1100... charge ordering, their results for $V_c$ are
completely consistent with ours in Section \ref{vcrit}. They have also
shown that dimerization increases $V_c$.

\section{Acknowledgments}

S.M. acknowledges support from the NSF under NSF-DMR-0101659.
D.K.C. acknowledges support from the NSF under NSF-DMR-97-12765.
Numerical calculations were done in part at the NCSA. 
We thank F. Zamborszky, S. Brown, and M. Meneghetti for useful
discussions.

\end{document}